\pdfoutput=1
\documentclass[aps,prl,floatfix,superscriptaddress,reprint,nolongbibliography]{revtex4-2}
\usepackage{eurosym}
\usepackage{amsbsy}
\usepackage{latexsym,epsfig,graphicx}
\usepackage{dcolumn}
\usepackage{subfigure}
\usepackage{comment}
\usepackage{color}
\usepackage{bm}
\usepackage{mathrsfs}
\usepackage{amssymb}
\usepackage{amsfonts}
\usepackage{amsmath}
\usepackage{xspace}
\usepackage{epstopdf}
\usepackage{tabularx}
\usepackage{longtable}
\usepackage[colorlinks=true, letterpaper=true, pdfstartview=FitV, linkcolor=red, citecolor=blue, urlcolor=blue]{hyperref}
\usepackage[normalem]{ulem}
\usepackage[version=4]{mhchem}
\usepackage{amsthm}
\usepackage[resetlabels]{multibib}
\newcites{SM}{SM-Literature}

\newtheorem{thm}{Theorem S\ignorespaces}

\begin{document}
\title{Graph Morphology of Non-Hermitian Bands}
\author{Yuncheng Xiong}
\affiliation{Beijing National Laboratory for Condensed Matter Physics, Institute of Physics, Chinese Academy of Sciences, Beijing 100190, China}
\author{Haiping Hu}
\thanks{Corresponding author: hhu@iphy.ac.cn}
\affiliation{Beijing National Laboratory for Condensed Matter Physics, Institute of Physics, Chinese Academy of Sciences, Beijing 100190, China}
\affiliation{School of Physical Sciences, University of Chinese Academy of Sciences, Beijing 100049, China}
\begin{abstract}
Non-Hermitian systems exhibit diverse graph patterns of energy spectra under open boundary conditions. Here we present an algebraic framework to comprehensively characterize the spectral geometry and graph topology of non-Bloch bands. Using a locally defined potential function, we unravel the spectral-collapse mechanism from Bloch to non-Bloch bands, delicately placing the spectral graph at the troughs of the potential landscape. The potential formalism deduces non-Bloch band theory and generates the density of states via Poisson equation. We further investigate the Euler-graph topology by classifying spectral vertices based on their multiplicities and projections onto the generalized Brillouin zone. Through concrete models, we identify three elementary graph-topology transitions (UVY, PT-like, and self-crossing), accompanied by the emergence of singularities in the generalized Brillouin zone. Lastly, we unveil how to generally account for isolated edge states outside the spectral graph. Our work lays the cornerstone for exploring the versatile spectral geometry and graph topology of non-Hermitian non-Bloch bands. 
\end{abstract}
\maketitle

{\color{blue}\textit{Introduction.---}}Non-Hermiticity emerges as a significant factor in a wide range of classical wave systems and open quantum systems, giving rise to various peculiar properties and applications \cite{coll1,coll4,coll6,colladd3,nhreview,nhreview2,nhreview3}. A distinctive feature of non-Hermitian systems is the non-Hermitian skin effect (NHSE) \cite{nhse1,nhse2,nhse3,nhse4,nhse5,nhse6,nhse7,nhse8,nhse9,nhse10,nhse11,nhsereview}, where an extensive number of eigenstates are localized at system boundaries \cite{nhsee1,nhsee2,nhsee3,nhsee4,zhuxueyi2020,liangqian2022}. To accommodate the presence of skin modes, the notion of generalized Brillouin zone (GBZ) \cite{nhse1} has been introduced by extending the Bloch wave vector to the complex domain. Depending on the boundary conditions, the energy spectra either manifest as closed loops with nontrivial spectral windings, representing Bloch bands under periodic boundary conditions (PBC), or they adopt open arcs (i.e., non-Bloch bands) on the complex-energy plane under open boundary conditions (OBC).

A synopsis of earlier non-Hermitian band theory involves symmetry classifications of eigenenergy bands based on point or line gaps \cite{nhclass1,nhclass2,nhclass3,nhclass4,nhclass5}. A homotopy perspective further distinguishes separable bands by their eigenenergy braidings \cite{homotopy1,homotopy2,nhknot,epknot,wangkai2021,huhaiping2022,yuyefei2022,caomm2023,zhangqicheng2023a,zhangqicheng2023b,wuyang2023}. These classifications, however, are exclusive to Bloch bands. Under OBC, the arc-shaped non-Bloch bands display a plethora of intricate patterns, forming planar graphs on the complex plane \cite{leechgraph,graph2} which are linked to the algebraic properties of characteristic polynomials (ChP). The primary focuses of the non-Bloch band theory \cite{nhse1,nhse2,nhse3,nhse4} have been on producing the continuum of non-Bloch bands and restoring the bulk-edge correspondence via the GBZ. Yet the intricate connections between these two types of spectral patterns (i.e., loops vs graphs), and the physical mechanism governing their transformations remain enigmatic. Broadly speaking, the intriguing graph geometry goes beyond the scope of conventional topological invariants like $\mathbb{Z}$, $\mathbb{Z}_2$ or spectral windings \cite{shen2018,nhse5,nhse6,yangzhesen}. It represents uncharted band topology which may yield novel non-Bloch symmetries \cite{nbpt1,nbpt2} and spectral transitions between distinct graph patterns relevant for anomalous responses \cite{leechgraph,qinfang2023}. To date, the non-Bloch bands with respect to their Euler-graph morphology have largely been unexplored, and a systematic classification of their spectral transitions is also lacking.

In this work, we present an algebraic framework to comprehensively characterize the graph geometry and topology of one-dimensional non-Hermitian bands. By incorporating a local potential function $\Phi(E)$, we unveil the electrostatic mechanism of spectral collapse and reproduce the non-Bloch band theory and the GBZ. Notably, the potential function is harmonic on the complex plane, except at locations coinciding with spectral graphs, aligning precisely with the troughs of the potential landscape. Furthermore, we delve into the Euler-graph topology and systematically classify the spectral vertices according to their multiplicities and projections onto the GBZ. We demonstrate three elementary graph transitions, as well as the appearance of singularities within the GBZ, and address the treatment of isolated edge modes that exist beyond the continuum of non-Bloch bands. 

\begin{figure}[!t]
	\includegraphics[width=3.3in]{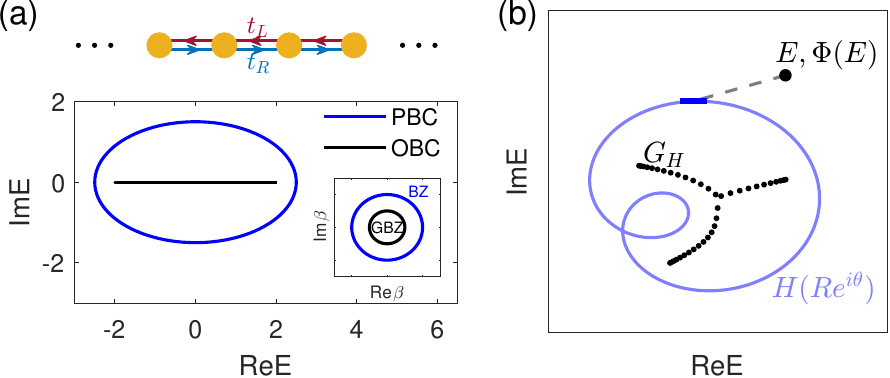}
	\caption{(a) (top) Schematics of the Hatano-Nelson model with asymmetric hoppings $t_R$ and $t_L$. (bottom) Energy spectra under PBC (blue) and OBC (black), respectively. $t_L=2, t_R=0.5$. Inset: loci of the Bloch factor $\beta=e^{ik}$ on the complex plane. (b) Sketch of the electrostatic analogy. By assigning a charge of $1/N$ ($N$ is the system size) to each point on the OBC spectra (black), the Coulomb potential $\Phi(E)$ felt at $E$ equals the electrostatic potential induced by the charged spectral loop described by $H(R e^{i\theta})$ with $\theta\in[0,2\pi]$. The factor $R$ is chosen such that $E$ is outside the loop.}\label{fig1}
\end{figure}

{\color{blue}\textit{Spectral collapse.---}}Let us first recap the non-Bloch band theory via the simplest Hatano-Nelson model \cite{hatano1996}. The Hamiltonian $H = \sum_j t_L c^\dagger_{j}c_{j+1} + t_R c^\dagger_{j+1}c_j$ consists of nearest hoppings to the left with strength $t_L > 0$ and to the right with $t_R > 0$, over a total of $N$ lattice sites. As shown in Fig. \ref{fig1}(a), the energy spectra form a closed oval under PBC while residing at the line connecting the two foci of the oval under OBC, $E\in[-2\sqrt{t_R t_L},2\sqrt{t_R t_L}]$. Consequently, all $N$ eigenstates are localized at either the left or right boundary depending on the ratio $t_L/t_R$. In momentum space, the Hamiltonian takes $H(k) = t_L e^{ik}+t_R e^{-ik}$. By substituting $\beta = e^{ik}$, it is easy to see that the eigenstates under OBC are skin modes associated with complex Bloch vectors, i.e., $\beta_m =  \sqrt{t_R/t_L}e^{\pm \frac{i 2\pi m}{N+1}}$, with $m$ the level index. In the continuum limit $N\rightarrow\infty$, all $\beta_m$'s form a circle of radius $\sqrt{t_R/t_L}$, which is the GBZ of the system.

The elegant spectral collapse into the oval's central axis is no accident. For a generic non-Hermitian Hamiltonian $H(k)$, the OBC energy spectra are determined by the bi-variate ChP ($\beta=e^{ik}$),
\begin{eqnarray}
	f(E,\beta)=\det(H(\beta)-E)=\sum_{j=-p}^{q}f_j(E) \beta^j.
\end{eqnarray}
For a given $E$, we order the solutions of $f(E,\beta)=0$ as $|\beta_1(E)|\leq|\beta_2(E)|\leq\cdots\leq|\beta_{p+q}(E)|$. In the continuum limit, the OBC spectra (except for a finite number of isolated edge states, if any) reside within the PBC spectral loop and form some Euler graph, denoted as $G_H$. Any point on the graph satisfies the condition $|\beta_p(E)|=|\beta_{p+1}(E)|$. The locus of $\beta$ on the complex plane obeying this condition gives the GBZ and traces a closed curve enclosing the origin.

The spectral collapse from PBC to OBC necessitates the point gap. Utilizing the so-called rescaled spectra $Sp(R)$ \cite{nhse6,hhp2d} which are the regions enclosed by the spectral loop of $H(R e^{i\theta})$ with $\theta\in[0,2\pi]$. The graph emerges as the intersection of these rescaled spectra with all feasible rescaling factors $R$:
\begin{eqnarray}\label{intersection}
	G_H=\mathop{\bigcap}\limits_{R\in(0,\infty)} Sp(R).
\end{eqnarray}
$G_H$ thus obtained is constituted by a collection of arcs \cite{bottcher2005,exceptions} and connected \cite{ullman1967}. Algebraically, the collapse process is captured by an electrostatic analogy \cite{wz2dnhse} which assigns a charge $1/N$ to each eigenvalue $E_n$ ($n=1,2,...,N$) of the OBC Hamiltonian. The Coulomb potential at $E\notin G_H$ is $\Phi(E)=\frac{1}{N}\sum_{n}\text{log}|E_n-E|$. According to Szeg\"o's limit theorem \cite{szego1,szego2}, the potential $\Phi(E)$ in the continuum limit equals the integral: $\Phi(E)=\frac{1}{2\pi}\int_0^{2\pi}\text{log}|\text{det}(H(R e^{i\theta})-E)|d\theta$ with $|\beta_{p}(E)|<R<|\beta_{p+1}(E)|$. Physically, the integral can be interpreted as the Coulomb potential due to the charged spectral loop $H(Re^{i\theta})$ with $\theta\in[0,2\pi]$, as sketched in Fig. \ref{fig1}(b). For any point $E\notin G_H$, the rescaling factor $R$ always exists. 

{\color{blue}\textit{Spectral graphs.---}}The potential function $\Phi(E)$ is well defined on the complement of the graph $G_H$. Notably, it can be represented  in a compact form \cite{SM}:
\begin{eqnarray}
	\Phi(E)=\log |f_q(E)|+\sum_{j=p+1}^{p+q}\log |\beta_j(E)|.
	\label{potential_z}
\end{eqnarray}
Thus, Eq. (\ref{potential_z}) can be straightforwardly continued to the whole complex plane without ambiguity as long as the $q$ roots of the largest moduli in the sum are chosen. The potential $\Phi(E)$ is harmonic outside the graph $G_H$, i.e., $\nabla^2_E\Phi(E)=0$ for $E\notin G_H$. In the distributional sense, the density of states (DOS) in the continuum limit is $ \rho(E)=\lim_{N\rightarrow\infty}\frac{1}{N}\sum_n\delta_{E,E_n}$. The electrostatic analogy implies that the DOS satisfies the Poisson equation
\begin{eqnarray}\label{poeq}
	\rho(E)=\frac{1}{2\pi}\nabla^2_E \Phi(E).
\end{eqnarray}
Clearly, a nonzero DOS can only be obtained when $E\in G_H$ or $|\beta_p(E)|=|\beta_{p+1}(E)|$, which is exactly the GBZ condition. On the complex-energy plane, the graph $G_H$ resides at the troughs of the potential landscape. The significant benefit of the potential formalism lies in its ability to derive the spectral graph and DOS without solving the GBZ or large OBC Hamiltonians.

\begin{figure}[!t]
	\includegraphics[width=3.3in]{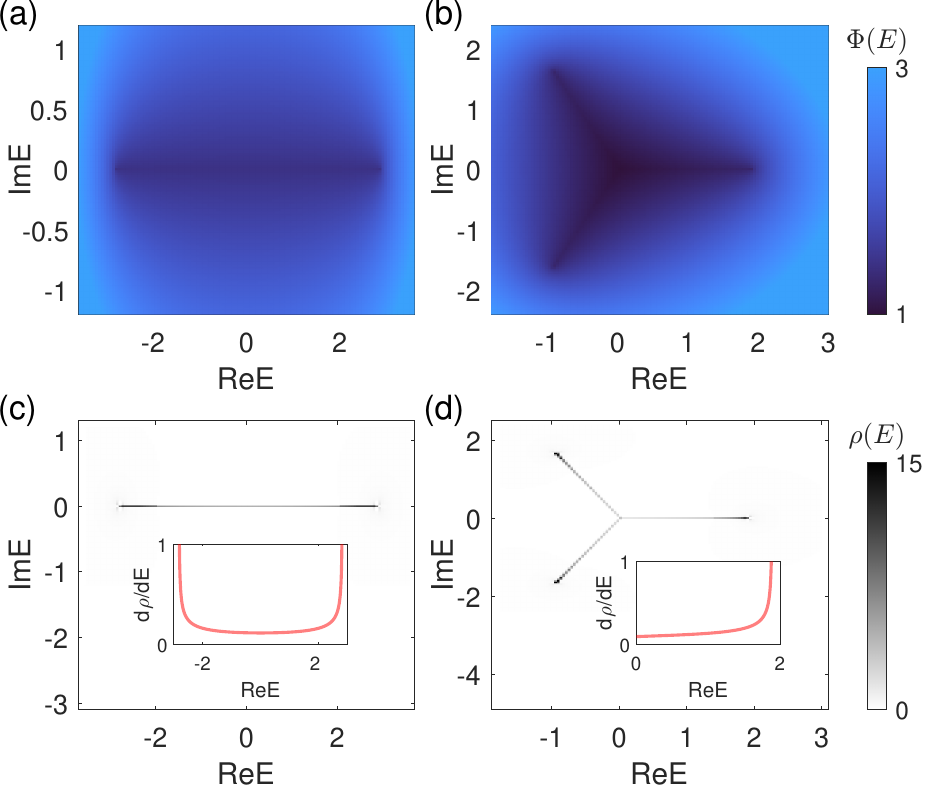}
	\caption{\label{fig2} Potential landscape on the complex plane associated with Hamiltonian (a) $H(\beta)=t_L\beta+t_R\beta^{-1}$ (the Hatano-Nelson model) and (b) $H(\beta)=\beta^2+\beta^{-1}$. (c)(d) The DOS $\rho(E)$ after taking the Laplacian of the potential functions. Inset: $\frac{d\rho}{dE}$, the DOS with respect to the arc length of the spectral graph. In the inset of (d), the branch on the real axis of the graph is selected. For (a)(c), $t_L=2, t_R=1$.}
\end{figure}

We illustrate the potential description with two simple examples. (\textit{i}) The Hatano-Nelson model. The potential is $\Phi(E)=\frac{1}{2}\log\max(|E\pm\sqrt{E^2-4t_R t_L}|)$. Figs. \ref{fig2}(a)(c) plot respectively the potential landscape and the DOS after taking the Laplacian on the complex plane. The DOS with respect to arc length \cite{SM} is $\frac{d\rho}{dE}=\frac{1}{\pi}\frac{1}{\sqrt{4t_L t_R-E^2}}$, which is divergent at the two endpoints. (\textit{ii}) $H(\beta)=\beta^2+\beta^{-1}$. The potential landscape is shown in Fig. \ref{fig2}(b). The OBC spectra form a three-fold symmetric fan with three branches: $G_H=\cup_{m} e^{i\frac{2m\pi}{3}}[0,3/\sqrt[3]{4}]$ ($m=0,1,2$), joining at the junction point $E=0$. The DOS with respect to the arc length \cite{SM} is $\frac{d\rho}{dE}=\frac{\sqrt{3}}{12\pi}\frac{1}{C_+^2+C_-^2-C_+ C_-}\Big(\frac{1}{4}-\frac{|E|^3}{27}\Big)^{-1/2}$, with $C_\pm=\Big(-\frac{1}{2}\pm\sqrt{\frac{1}{4}-\frac{|E|^3}{27}}\Big)^{1/3}$, as shown in Fig. \ref{fig2}(d). For more complicated cases, the potential function, spectral graph, and DOS can be solved numerically. 

\begin{figure}[!t]
	\includegraphics[width=3.3in]{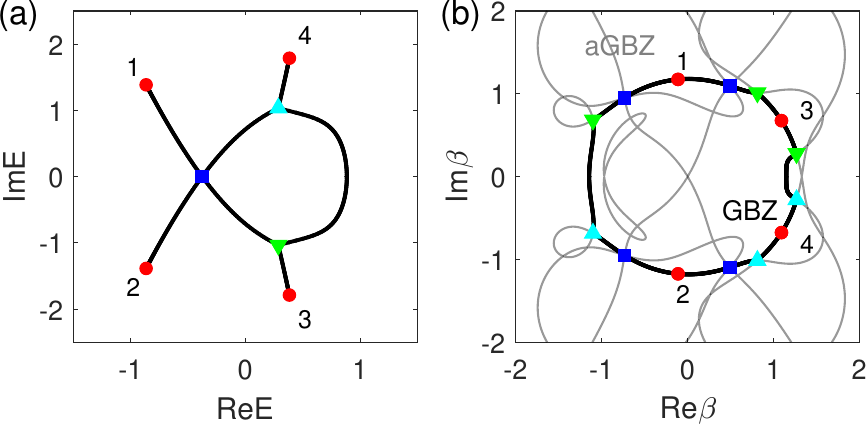}
	\caption{\label{fig3} Euler-graph topology associated with Hamiltonian $H(\beta)=\beta^{-3}+0.99\beta^{-2}+0.1\beta^2-0.44\beta^3$. (a) The graph $G(7,7)$ formed by the continuum of non-Bloch band, with four endpoints (red dots), one 4-vertex (blue square), and two 3-vertices (green and cyan triangles). (b) The GBZ (black) and auxiliary GBZ curves (gray) on the $\beta$-plane. The projections of the endpoints and junction points in (a) are displayed in the same markers/colors.}
\end{figure}

{\color{blue}\textit{Euler-graph topology.---}}The topology of the graph $G_H$ is specified by the number of vertices ($n_V$), edges ($n_E$) and faces ($n_F$). They satisfy the Euler formula \cite{multicomponent}:
\begin{eqnarray}\label{euler}
	n_V-n_E+n_F=2.
\end{eqnarray}
For clarity, we also label a graph by $G(n_V,n_E)$ hereafter. Usually, the information of a finite set of points on $G_H$, e.g., the endpoints and the junction points with their multiplicities, is sufficient to determine the graph topology. These special points can be obtained via fairly simple analytical methods. To this end, we classify the points on the graph by their local geometric structures. Let us take a small circle enclosing $E\in G_H$ and dub $E$ an $n$-vertex if there are $n$ arcs emanating from it and intersecting with the circle. Thus the points on the edges (interior of the arcs) of $G_H$ are 2-vertices and the endpoints are 1-vertices. The junctions are $n$-vertices with $n\geq 3$ (e.g., $E=0$ in Fig. \ref{fig2}(d)). Fig. \ref{fig3} plots the spectral graph and its associated GBZ for Hamiltonian $H(\beta)=\beta^{-3}+0.99\beta^{-2}+0.1\beta^2-0.44\beta^3$. The graph $G_H=G(7,7)$ contains two 3-vertices, one 4-vertex, and four endpoints due to the existence of a closed loop.

\begin{figure}[!t]
	\includegraphics[width=3.3in]{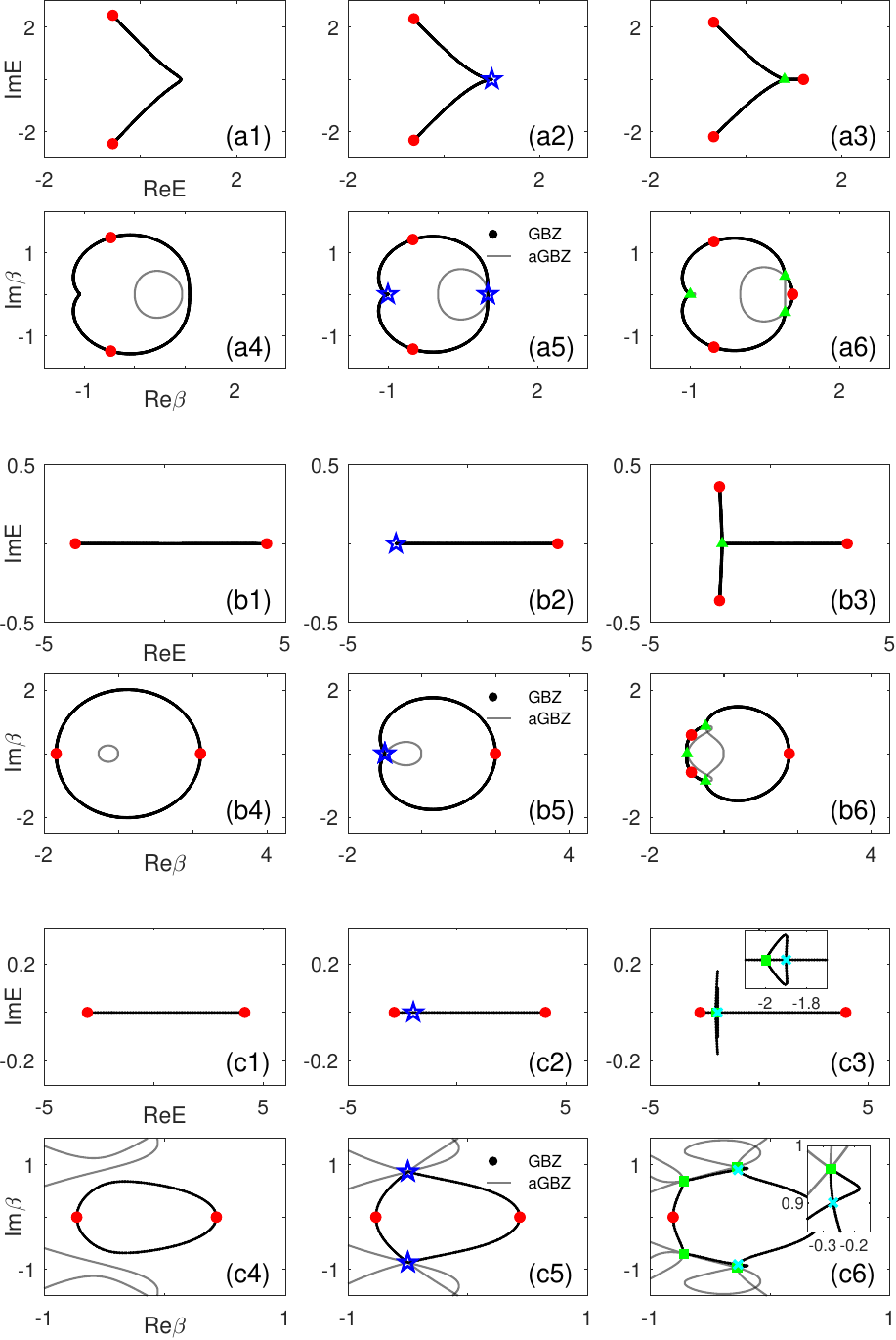}
	\caption{\label{fig4} Three elementary spectral-graph transitions. (a1-a6): UVY transition. (b1-b6): PT-like transition. (c1-c6): self-crossing transition. For each type, the first and second row denotes the evolutions of the spectral graph and its corresponding GBZ, respectively. The endpoints, junction points, and the spectral singularities at the transitions are marked in red, green, and blue, respectively. In (c3), the spurious 4-vertex is marked in cyan. For (a1-a6), from left to right, $\alpha=-1.2,-1,-0.8$. For (b1-b6), from left to right, $\alpha=4,3,2$. For (c1-c6), from left to right, $\alpha=3.2,3,2.8$.} 
\end{figure}

Different types of vertices have distinct projections onto the GBZ. As per the potential theory, for any point $E\in G_H$, the ``middle" two solutions of the ChP share an equal modulus $|\beta_{p}(E)|=|\beta_{p+1}(E)|$. When one travels around the GBZ, any arc within the graph $G_H$ must be traversed in both directions. Consequently, the projections of a normal 2-vertex onto GBZ manifest as two separate points of the same modulus. The endpoint corresponds to the scenario of degenerate roots $\beta_{p}(E)=\beta_{p+1}(E)$, obeying:
\begin{eqnarray}\label{res1}
	f_{Res}\equiv\textrm{Res}_\beta[\beta^p f(E,\beta),\partial_\beta(\beta^p f(E,\beta))]=0,
\end{eqnarray}
with $\textrm{Res}$ the resultant function \cite{SM,yangzhesen,resmath1,resmath2,gelfand1994} to discriminate degenerate solutions of the ChP. The junction point, or $n$-vertex ($n\geq 3$) has $n$ separate projections onto the GBZ. Analytically, the junction point $E$ has $n$ successive $\beta$-solutions of the same modulus \cite{wuprb}. Each projection lies at the crossing between the algebraic curves of the auxiliary GBZ \cite{nhse7}, 
e.g., the projections in Fig. \ref{fig3}(b) of the 3- or 4-vertex. It is important to note that there may exist some spurious junctions on the graph arising from spectral self-crossings, as exemplified in Fig. \ref{fig4}(c3). In such instances, the GBZ loop displays self-intersections and the fake junction should be treated as a normal 2-vertex on the spectral arc.

{\color{blue}\textit{Graph transitions.---}}The graph geometry enables uncharted spectral-graph transitions without any Hermitian analog. Unlike the band-touching-induced topological transitions responsible for the appearance of edge states, graph transitions may occur even in single-band non-Hermitian systems. Guided by the Euler formula, these transitions involve changes in the counts of vertices, edges, or faces adhering to the condition $\delta n_V - \delta n_E + \delta n_F = 0$. Given the diversity of spectral graphs, there exists an infinite array of graph transitions, which can be further broken down into more elementary ones. In the following, we pinpoint three fundamental graph transitions with concrete examples.

(i) UVY transition. Model: $H(\beta)=\beta+\alpha/\beta+1/\beta^2$ with a tunable parameter $\alpha$. As shown in Figs. \ref{fig4}(a1-a3), the number of endpoints changes by one, $G(2,1)\rightarrow G(4,3)$. Initially, at $\alpha=-1.2$, there is a smooth U-turn in the arc. As $\alpha$ increases, the U-turn becomes narrower and sharper to develop a cusp at $\alpha=-1$. After that, a new endpoint emerges, and the graph becomes Y-shaped with three arcs joined at a 3-vertex. Figs. \ref{fig4}(a4-a6) plot the GBZ. At the transition, two aGBZ curves touch.

(ii) PT-like transition. Model: same as above, but with $\alpha$ varying from $4$ to $2$. As shown in Figs. \ref{fig4}(b1-b3), the number of vertices and edges change by $\delta n_V = \delta n_E = 2$ with $G(2,1)\rightarrow G(4,3)$, accompanied by the appearance of a $3$-vertex. Note that the two new branches may also emerge from the interior of the spectral arc \cite{nbptwz}, rather than at the endpoint. This is a generalization of non-Bloch PT-breaking, where the spectra transition from real to complex, or vice versa. Similar to the former case, two aGBZ curves touch at the transition with $\alpha=3$, as depicted in Fig. \ref{fig4}(b5).

(iii) Self-crossing transition. Model: $H(\beta)=\beta^3+2\beta^2+\alpha \beta+1/\beta$, with $\alpha$ varying from $3.2$ to $2.8$. In this case, the number of endpoints stays unchanged, but new junction points and spectral loops emerge. As depicted in Figs. \ref{fig4}(c1-c3), the initial graph $G(2,1)$ transforms into an ``airplane" graph $G(3,3)$. Due to the appearance of an additional spectral loop, $\delta n_F=1$. Note that the 4-vertex on the right side (cyan dot in Fig. \ref{fig4}(c3)) is spurious as it maps onto two (rather than four) solutions on the GBZ [Fig. \ref{fig4}(c6)]. Nonetheless, when traversing the GBZ, it is passed through four times. This spectral loop spans from the intermediate endpoint at $E=-2$ with $\alpha=3$. Still, the aGBZ curves touch at the transition, resulting in the GBZ's self-crossings afterward.

We remark that all three types of transitions involve the emergent singularities in the GBZ. In the UVY/PT-like transitions, the number of endpoints changes, which can be identified by solving all the endpoints via Eq. (\ref{res1}). While for the PT-like/self-crossing transitions, there exist two coincident endpoints as shown in Fig. \ref{fig4}(b2)(c2). Similar to the criterion of identifying the non-Bloch PT-transition \cite{nbptwz}, the spectral transition satisfies the condition:
\begin{align}
	\text{Res}_E\Big[f_{Res}(E,\alpha),\partial_E f_{Res}(E,\alpha)\Big]=0.
	\label{eq:transition_point}
\end{align}

{\color{blue}\textit{Isolated edge states.---}}In realistic systems, isolated topological edge states may exist outside the graph. Consider a generic $m$-band ($m\geq 2$) Hamiltonian $H(\beta)=\sum_{j=-s}^{t}h_j \beta^j$ with $h_j$ being $m\times m$ matrices and $s(t)$ the hopping range to the right (left). To accommodate the edge state $E\notin G_H$, we define an $ms\times ms$ matrix $M_{edge}$, with its $\mu$th row and $\nu$-th column ($\mu,\nu=1,2,\cdots,s$) an $m\times m$ block, given by
\begin{eqnarray}
	[M_{edge}]_{\mu\nu}=\frac{1}{2\pi i}\int_{\Gamma} \beta^{\mu-\nu}(H(\beta)-E~I_{m\times m})^{-1} \frac{d\beta}{\beta}.
	\label{eq:C0}
\end{eqnarray}
Here $I_{m\times m}$ is the identity matrix. The line integral is defined entry-wise, tracing a counterclockwise closed curve $\Gamma$ which encloses the origin and the first $p$ $\beta$-solutions of $f(E,\beta)$. Isolated edge modes are identified by the condition $\det M_{edge}=0$ \cite{edgemode}. For instance in the non-reciprocal SSH model \cite{nhse1}: $H(k)=(t_1+t_2\cos k)\sigma_x+(t_2\sin k+i\frac{\gamma}{2})\sigma_y$, $m=2$, $s=p=1$. It is easy to show $\det M_{edge}\neq 0$ if $E\neq 0$. The admissible edge states manifest as zero modes when $|t_1^2-\gamma^2/4|<t_2^2$ \cite{SM}. It is the topological regime indexed by a nonzero winding number \cite{nhse1} over the GBZ. The edge matrix is powerful in analyzing isolated edge states, including non-zero edge modes \cite{SM} without the need of bulk-invariant calculations.

{\color{blue}\textit{Conclusion and discussion.---}}To conclude, our study systematically delves into the geometric and topological aspects of spectral graphs of non-Hermitian non-Bloch bands, within a purely algebraic framework. Facilitated by the local form of the potential function, we unveil the electrostatic mechanism of spectral collapse, reproduce the non-Bloch band theory as well as the GBZ, and extract the spectral graph from the potential landscape. By analyzing the Euler-graph topology, we further identify three elementary spectral transitions beyond the gap-closure paradigm and address how to account for isolated edge states outside the continuum of non-Bloch bands.

The local potential function plays a vital role in bridging the graph topology of non-Hermitian non-Bloch bands with the underlying algebraic structures of the ChP. While the spectral graph is immediately derived from the potential landscape or the intersection in Eq. (\ref{intersection}), it can also be obtained by locating self-crossing points on the rescaled spectral loops \cite{wuprb}, or through the auxiliary GBZ method \cite{nhse7}. In contrast to the traditional wisdom of computing energy spectra for a given Hamiltonian, a reverse band-engineering strategy can be employed \cite{lee_sci} to design tailored tight-binding Hamiltonians for a specific spectral graph. Notably, the potential landscape in our work differs from the inverse skin-depth landscape therein. Lastly, the intricate spectral graphs and their graph-topology transitions could be readily explored in platforms like metamaterials \cite{nhsee1}, optics/photonics \cite{nhsee3,zhuxueyi2020}, ultracold atoms \cite{liangqian2022}, and electrical networks \cite{nhsee2,nhsee4}.
\begin{acknowledgments}
	This work is supported by the National Key Research and Development Program of China (Grant No. 2022YFA1405800) and the start-up grant of IOP-CAS.
\end{acknowledgments}

\bibliographystyle{revisedapsrev4-2}
\bibliography{nhgraph}

\begin{thebibliography}{65}
\bibitem [{mul()}]{multicomponent}%
  \BibitemOpen
\bibfield  {journal} {  }\href@noop {} {\bibinfo  {journal} {If the spectral
  graph has $n_C$ components, the right side of the Euler formula should be
  modified to be $1+n_C$}\ }\BibitemShut {NoStop}%
\end{thebibliography}

\clearpage
\appendix
\renewcommand{\thefigure}{S\arabic{figure}}
\renewcommand{\theequation}{S\arabic{equation}}
\setcounter{figure}{0}
\pagebreak
\widetext
\begin{center}
	\textbf{\large  Supplemental Material}
\end{center}

This supplemental material provides additional details on

(I) The derivation of the local spectral potential;

(II) The linear density of state of the Hatano-Nelson model and the ``fan" model;                                                                                                                                                                                                                                                       

(III) The resultant function;

(IV) Calculations of isolated topological edge modes outside the spectral graph.

\section{(I) Derivation of the local potential function}\label{seci}
In this section, we rigorously derive the local potential function (Eq. (3) in the main text). We analytically continue the non-Hermitian Hamiltonian $H(k\rightarrow\beta=e^{ik})$ and examine the characteristic polynomial (ChP): 
\begin{eqnarray}\label{chp}
	f(E,\beta)=\det(H(\beta)-E)=\sum_{j=-p}^{q}f_j(E) \beta^j,
\end{eqnarray}
with two complex variables. The energy spectra under periodic boundary conditions (PBC) are given by the roots of the ChP $\sigma_{PBC}=\{E, f(E,\beta)=0\}$ with $|\beta|=1$. Generally, PBC spectra trace closed loops on the complex plane, while the energy spectra under open boundary conditions (OBC) are enclosed within these loops, except for some finite number of topological edge states. Now, let us explore the deformation of the PBC spectral loop formed by the eigenvalues of $H(Re^{i\theta})$ as $\theta$ varies from $0$ to $2\pi$. $R=1$ corresponds to the PBC spectra. It turns out that for all rescaling factor $R\in(0,\infty)$, the OBC spectra remain enclosed within these deformed loops. 

The OBC spectra are linked to the above deformed spectra through Szeg\"o's strong limit theorem \cite{szego1,szego2}. For an $N\times N$ OBC Hamiltonian $H$ (Teoplitz matrix), in the $N\rightarrow\infty$ limit (continuum limit), its eigenvalues $E_1,E_2,...,E_N$ would form a planar graph (denoted as $G_H$), composed of open arcs on the complex plane. Assigning each eigenvalue a charge $1/N$, the Coulomb potential at $E\notin G_H$ induced by these charged eigenvalues is 
\begin{align}
	\Phi(E)=\frac{1}{N}\sum_{n}\text{log}|E_n-E|.
	\label{eqSM:potential_obc}
\end{align}
The  Szeg\"o's strong limit theorem states that up to $O(1/N)$ correction, this potential is expressed by the integral:
\begin{eqnarray}
	\Phi(E)=\frac{1}{2\pi}\int_0^{2\pi}\text{log}|\text{det}(H(Re^{i\theta})-E)|d\theta=\frac{1}{2\pi}\int_0^{2\pi}\text{log }|f(E,Re^{i\theta})|d\theta.
	\label{eqSM:potential_rpbc}
\end{eqnarray}
Physically, the integral represents the electrostatic potential experienced by a test charge at location $E$ due to the eigenvalues on the deformed loop $H(Re^{i\theta})$ with $\theta\in[0,2\pi]$. Here $R$ is chosen such that $|\beta_p|<R<|\beta_{p+1}|$.

To compute the integral in Eq. (\ref{eqSM:potential_rpbc}), we label the $\beta$-solutions of $f(E,\beta)=0$ as $|\beta_1|\leq\cdots\leq|\beta_p|<|\beta_{p+1}|\leq\cdots\leq|\beta_{p+q}|$. Since $E\notin G_H$, the inequality $|\beta_p|<|\beta_{p+1}|$ is guaranteed. The ChP can be expressed using these $\beta$-solutions as $f(E,Re^{i\theta})=f_q(Re^{i\theta})^{-p}\prod_{j=1}^{p+q}(Re^{i\theta}-\beta_j)$. The potential function is then expanded as:
\begin{align}
	\Phi(E)=\log|f_q|+\frac{1}{2\pi} \text{Re}\int_0^{2\pi}d\theta \left[\sum_{j=1}^{p+q}\log(Re^{i\theta}-\beta_j)-p\log(Re^{i\theta})\right],
	\label{phi2}
\end{align} 
where Re denotes the real part and Re$(\log(x))=\log|x|$ is used. Utilizing the Taylor expansion, we obtain
\begin{align*}
	&\log(Re^{i\theta}-\beta_j)=\log(Re^{i\theta})+\sum_{l=1}^{\infty}\frac{(-1)^l}{l}(\beta_j/R)^l e^{-il\theta}, \quad 1\leq j\leq p; \\
	&\log(Re^{i\theta}-\beta_j)=\log(-\beta_j)+\sum_{l=1}^{\infty}\frac{(-1)^l}{l}(R/\beta_j)^l e^{il\theta}, \quad p+1\leq j\leq p+q.
\end{align*}
The integration over $\theta\in[0,2\pi]$ in Eq. (\ref{phi2}) vanishes for all terms with $l\geq 1$. Consequently, the potential function simplifies to the desired form:
\begin{eqnarray}
	\Phi(E)=\log |f_q(E)|+\sum_{j=p+1}^{p+q}\log |\beta_j(E)|.
	\label{eqSM:potential_local}
\end{eqnarray}
We note that the integral form of the potential function in Eq. (\ref{eqSM:potential_rpbc}) only applies to $E\notin G_H$. For $E\in G_H$, $|\beta_p|=|\beta_{p+1}|$, the integral becomes ill-defined. 
In contrast, the local form in Eq. (\ref{eqSM:potential_local}) eliminates this ambiguity, allowing the potential function to be extended to the entire complex plane. For any $E\in\mathbb{C}$, the $\beta$-solutions of $f(E,\beta)$ can be sorted as $|\beta_1|\leq\cdots\leq|\beta_p|\leq|\beta_{p+1}|\leq\cdots\leq|\beta_{p+q}|$. The order of solutions with the same modulus can be freely selected, as only the moduli are relevant in Eq. (\ref{eqSM:potential_local}). We then choose  the $q$ $\beta$-solutions with the largest moduli to construct the potential function by Eq. (\ref{eqSM:potential_local}). Importantly, the local potential function simplifies the computation of the continuum of non-Bloch bands through an electrostatic analogy (as discussed in the main text and also in Sec (II) below). This approach circumvents the notorious errors in numerical diagonalizations of large non-Hermitian Hamiltonians and the cumbersome integrations in Eq. (\ref{eqSM:potential_rpbc}).

\section{(II) The linear DOS of the Hatano-Nelson model and the ``fan" model}
\begin{figure}[ht!]
	\includegraphics[width=0.3\textwidth]{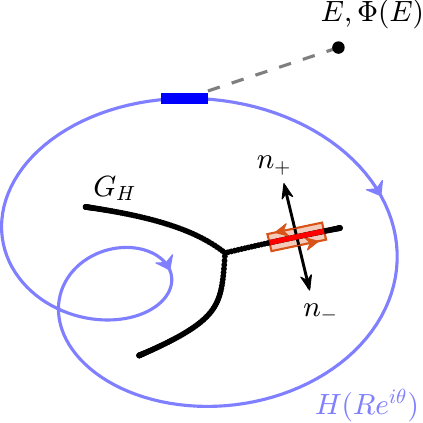}
	\caption{\label{figS1} Sketch of the spectral graph on the complex-energy plane. The rescaled spectra (light blue curve) described by $H(Re^{i\theta})$ trace an oriented closed loop with varying $\theta$ from $0$ to $2\pi$, while the energy spectra under OBC form a planar graph (black arcs). On the graph, $n_+$ and $n_-$ denote the opposing normal vectors to the line element (red segment) on the spectral arc. Green's theorem is applied within the orange area element with oriented boundaries.}
\end{figure}

The density of states (DOS) characterizes the arrangement of eigenvalues of a non-Hermitian Hamiltonian on the complex plane. In a distributional sense, the DOS in the continuum limit is expressed as:
\begin{eqnarray}
	\rho(E)=\lim_{N\rightarrow\infty}\frac{1}{N}\sum_n\delta_{E,E_n}.
	\label{eqSM:rho}
\end{eqnarray}
The electrostatic analogy tells that the DOS is linked to the spectral potential through the Poisson equation (up to a $\frac{1}{2\pi}$ factor):
\begin{align}
	\rho(E)=\frac{1}{2\pi}\Delta\Phi(E).
	\label{eqSM:laplacian}
\end{align}
This becomes apparent from Eqs. (\ref{eqSM:potential_obc})(\ref{eqSM:rho}) by recalling that $\frac{1}{2\pi}\text{log}|z|$ ($z\in\mathbb{C}$) is the Green's function of the Laplacian operator on the complex plane. Since the OBC spectra form a planar graph $G_H$, it is more relevant to consider the linear DOS, which is the DOS measured along the arc length of the spectral graph. Applying Green's theorem, the linear DOS is expressed as:
\begin{align}
	\frac{d\rho}{dE}=\frac{1}{2\pi}\big( \frac{\partial \Phi(E)}{\partial n_+}+\frac{\partial \Phi(E)}{\partial n_-} \big).
	\label{eqSM:lineDOS}
\end{align}
Here $dE$ denotes the line element on the graph, and $n_+$ and $n_-$ are the two opposing normal directions of the arc element, as sketched in Fig. (\ref{figS1}). In the following, we employ Eqs. (\ref{eqSM:potential_local})(\ref{eqSM:lineDOS}) to determine the linear DOS of two specific models that allow for analytical treatment.   

\subsection{(A) Hatano-Nelson model} 
The Hatano-Nelson model features nonreciprocal hoppings, described by Hamiltonian $H(\beta)=t_L \beta+t_R \beta^{-1}$. Its OBC spectra lie at the interval $G_H=[-2\sqrt{t_L t_R},2\sqrt{t_L t_R}]$ on the real line. The ChP is $f(E,\beta)=t_L \beta+t_R \beta^{-1}-E$, corresponding to $p=q=1$ in Eq. (\ref{chp}). For any $E\in\mathbb{C}$, the local potential is the maximum of the two $\beta$-solutions: 
\begin{align}
	\Phi(E)=|t_L| \text{ max}(\Big|\frac{E+\sqrt{E^2-4t_L t_R}}{2t_L}\Big|,\Big|\frac{E-\sqrt{E^2-4t_L t_R}}{2t_L}\Big|).
	\label{eqSM:potential_hn}
\end{align}
For $E\in G_{H}$, $|\beta_+|=|\beta_-|$. Away from the graph, $|\beta_+|>|\beta_-|$ or $|\beta_+|<|\beta_-|$ depending on the specific $E$. To obtain the linear DOS, we need the normal derivative of $\Phi(E)$ for $E\in G_H$. It is easy to check that for any $E\in G_H$ (not the endpoints), $|\beta_-(E+i\epsilon)|>|\beta_+(E+i\epsilon)|$ and $|\beta_+(E-i\epsilon)|>|\beta_-(E-i\epsilon)|$ with $\epsilon\to0^+$. Thus, the linear DOS is
\begin{align}
	\frac{d\rho}{dE}&=\lim_{\epsilon\to 0}\frac{\Phi(E+i\epsilon)-\Phi(E)}{|\epsilon|}+\frac{\Phi(E-i\epsilon)-\Phi(E)}{|\epsilon|} \nonumber\\
	&=|t_L|\lim_{\epsilon\to 0}\frac{|\beta_-(E+i\epsilon)|-|\beta_-(E)|}{|\epsilon|}+\frac{|\beta_+(E-i\epsilon)|-|\beta_+(E)|}{\epsilon} \nonumber\\
	&=\frac{1}{\pi}\frac{1}{\sqrt{4t_L t_R-E^2}}.
	\label{eqSM:dosA_HN}
\end{align}
The linear DOS diverges at the two endpoints of the spectral graph.

\subsection{(B) The ``fan" model} 
For the fan model with Hamiltonian $H(\beta)=\beta^2+\beta^{-1}$, the ChP is $f(E,\beta)=\beta^2+\beta^{-1}-E$. The spectral graph resembles a 3-fold symmetric fan, comprising three branches  $e^{i\frac{2\pi}{3}m}\xi$ with $\xi\in[0,3/\sqrt[3]{4}]$. Here $m=0,1,2$ labels the branches. For any $E\in\mathbb{C}$, the three $\beta$-solutions are:
\begin{align*}
	\beta_j(E)=e^{i\frac{2\pi}{3}j}C_+ + \frac{1}{3} E e^{-i\frac{2\pi}{3}j}C_+, \quad j=1,2,3.
\end{align*}
They simplify to $\beta_j(E)=e^{i\frac{2\pi}{3}j}C_+ + e^{i\frac{2\pi}{3}(m-j)}C_-$ ($j=1,2,3$) if $E\in G_H$, with $C_\pm=\sqrt[3]{-\frac{1}{2}\pm\sqrt{\frac{1}{4}-\frac{E^3}{27}}}$. For the $m$-branch of $G_H$, we choose the two opposing infinitesimal normal vectors as $\pm\epsilon_m=\pm i|\epsilon|e^{i\frac{2\pi}{3}m}$ with $|\epsilon|\to 0$. For instance, on the $m=1$ branch, one can verify that $|\beta_2(E+\epsilon_1)|>|\beta_3(E+\epsilon_1)|>|\beta_1(E+\epsilon_1)|$ and $|\beta_2(E-\epsilon_1)|>|\beta_1(E-\epsilon_1)|>|\beta_3(E-\epsilon_1)|$ with $|\epsilon|\to 0$. Thus,
\begin{align*}
	\frac{d\rho}{dE}=\frac{1}{2\pi}\lim_{|\epsilon|\to 0}\frac{|\beta_3(E+\epsilon_1)\beta_2(E+\epsilon_1)|-|\beta_3(E)\beta_2(E)|}{|\epsilon|}+\frac{|\beta_1(E-\epsilon_1)\beta_2(E-\epsilon_1)|-|\beta_1(E)\beta_2(E)|}{|\epsilon|},
\end{align*}
with $\epsilon_1=i|\epsilon|e^{i\frac{2\pi}{3}}$. Explicit expressions for the other two branches are obtained similarly:
\begin{align*}
	&\frac{d\rho}{dE}=\frac{1}{2\pi}\lim_{|\epsilon|\to 0}\frac{|\beta_1(E+\epsilon_0)\beta_3(E+\epsilon_0)|-|\beta_1(E)\beta_3(E)|}{|\epsilon|}+\frac{|\beta_2(E-\epsilon_0)\beta_3(E-\epsilon_0)|-|\beta_2(E)\beta_3(E)|}{|\epsilon|}, ~~m=0\\
	&\frac{d\rho}{dE}=\frac{1}{2\pi}\lim_{|\epsilon|\to 0}\frac{|\beta_2(E+\epsilon_2)\beta_1(E+\epsilon_2)|-|\beta_2(E)\beta_1(E)|}{|\epsilon|}+\frac{|\beta_3(E-\epsilon_2)\beta_1(E-\epsilon_2)|-|\beta_3(E)\beta_1(E)|}{|\epsilon|}, ~~m=2.
\end{align*}
A straightforward calculation yields that the linear DOS for the three branches is independent of $m$ and given by:
\begin{align}
	\frac{d\rho}{dE}=\frac{\sqrt{3}}{12\pi}\frac{1}{C_+^2+C_-^2-C_+ C_-}\Big(\frac{1}{4}-\frac{|E|^3}{27}\Big)^{-1/2}.
	\label{eqSM:dosA_z3}
\end{align}
The linear DOS depends on the modulus of $E$ and diverges at the endpoints of the graph.
\begin{figure}[!b]
	\includegraphics[width=0.65\textwidth]{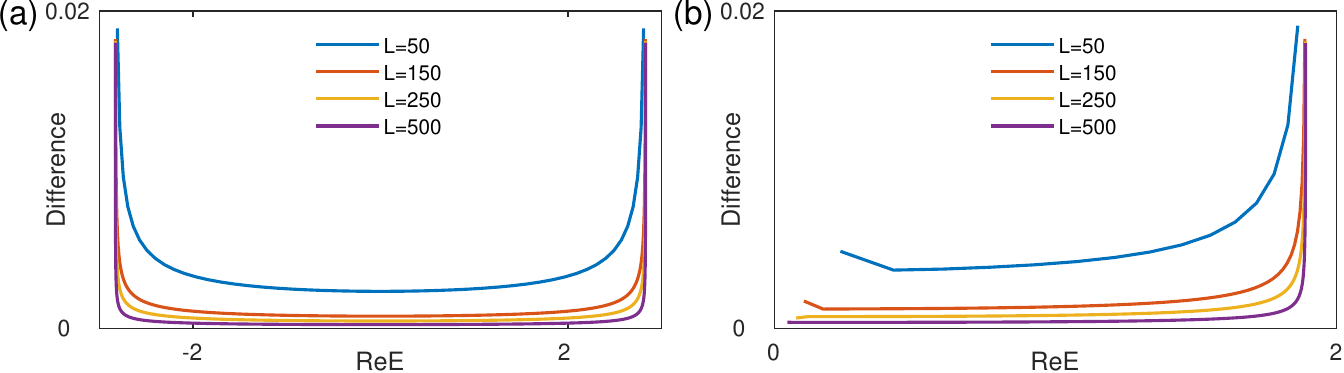}
	\caption{\label{figS2} The disparities between the analytical linear DOS and numerical results via exact diagonalization for (a) the Hatano-Nelson (HN) model and (b) the fan model. Four system sizes are taken $L=50, 150, 250$, and $500$. In (b), the linear DOS along the $m=0$ branch of the spectral graph is examined.}
\end{figure}

\subsection{(C) Comparison with numerical estimates}
We compare the linear DOS obtained from the potential formulation (Eqs. (\ref{eqSM:dosA_HN})(\ref{eqSM:dosA_z3})) with numerical results. For any eigenvalue $E$ (excluding endpoints of the OBC spectra) in the finite-size system, we take the two nearest eigenvalues, denoted as $E_a$ and $E_b$, positioned on opposite sides of $E$. The average spectral distance from $E$ to its nearest neighbors is $(|E-E_a|+|E_b-E|)/2$. The linear DOS is given by
\begin{align}
	[\frac{d\rho}{dE}]_{num}=\frac{1}{L}\frac{2}{|E-E_a|+|E_b-E|},
	\label{eqSM:dosN}
\end{align}
with $L$ the total number of sites. $[\frac{d\rho}{dE}]_{num}$ is automatically normalized to the number of bands. In Fig. \ref{figS2}(a)(b), we illustrate the discrepancies between numerical results and the analytical linear DOS in Eqs. (\ref{eqSM:dosA_HN})(\ref{eqSM:dosA_z3}) for the Hatano-Nelson model and the fan model, respectively. We consider four lattice lengths, namely $L=50, 150, 250, 500$, to estimate the linear DOS. The monotonic decrease in discrepancies with increasing system size is evident. The visible buckling at the endpoints is attributed to the divergence of the linear DOS there. Notably for the two models, a relatively small system size (e.g., the blue line with $L=50$ in Fig. \ref{figS2}) is enough to provide a good estimation for the linear DOS in the continuum limit. 

\section{(III) The resultant function}
In the main text, we employed the resultant function to identify the endpoints of the spectral graph and to determine Type-II and Type-III spectral-graph transitions. The resultant function serves as a versatile algebraic tool with various applications. Here we offer a concise introduction to the resultant of two polynomials, and its connection to the Sylvester matrix \cite{res1,res2,res3}. For two univariate polynomials $f(x)=\sum_{i=0}^n f_i x^i$ with zeros $\alpha_i, i=1,2,\cdots,n$, and $g(x)=\sum_{j=0}^{m} g_j x^j$ with zeros $\beta_j, j=1,2,\cdots,m$, their resultant is defined as
\begin{align*}
	\text{Res}_x[f,g]=(f_n)^m(g_m)^n \prod_{i,j}(\alpha_i-\beta_j).
\end{align*}
Here the subscript $x$ signifies that the resultant is computed with respect to the variable $x$. It is evident that the resultant can identify whether two polynomials share (at least one) common zeros.

\begin{thm}
	The resultant of $f(x)$ and $g(x)$ equals zero if and only if the two polynomials have a common zero.
\end{thm}

The resultant of two polynomials can be extended to multivariate scenarios, provided that all variables, except the specified one (i.e., the subscripted variable), are treated as parameters. Although computing the resultant may appear to necessitate knowledge of all the roots of the two polynomials, a more straightforward approach involves the Sylvester matrix, which is constructed from the coefficients of the polynomials as follows:
\begin{align*}
	Syl_x[f,g]=\begin{pmatrix}
		f_n & f_{n-1} & \cdots & f_2 & f_1 & 0 & \cdots \\
		0 & f_n & \cdots & f_3 & f_2 & f_1 & \cdots \\
		\vdots & \vdots & \vdots & \vdots & \vdots & \vdots & \vdots \\
		g_m & g_{m-1} & \cdots & g_2 & g_1 & 0 & \cdots \\
		0 & g_m & \cdots & g_3 & g_2 & g_1 & \cdots \\
		\vdots & \vdots & \vdots & \vdots & \vdots & \vdots & \vdots
	\end{pmatrix}.
\end{align*}
The Sylvester matrix $Syl_x$ is an $(m+n)\times(m+n)$ matrix with its first $m$ (last $n$) rows consisting of the coefficients of $f(x)$ ($g(x)$). Alternatively, the Sylvester matrix can be written entrywise as
\begin{align*}
	\big(Syl_x[f,g]\big)_{ij}=
	\begin{cases}
		f_{i-j+n}  &  0\leq j-i<n ~\&~ 0<i\leq m \\
		g_{i-j}  & 0<i-j\leq m ~\&~ m<i\leq m+n \\
		0 & \text{otherwise}
	\end{cases}.
\end{align*}

\begin{thm}
	The determinant of the Sylvester matrix constructed from the coefficients of the polynomials $f(x)$ and $g(x)$ is equal to the resultant of the two polynomials, i.e.,
	{\upshape
		\begin{align}
			\text{det}Syl_x[f,g]=\text{Res}_x[f,g].
		\end{align}
	}
\end{thm}

In the following, we elaborate on how the resultant can be used to determine the endpoints and topological transitions of the spectral graph. Given the ChP $f(E,\beta)=\sum_{j=-p}^{q}f_j(E)\beta^j$, the endpoints correspond to the stringent constraint that the two zeros $\beta_p=\beta_{p+1}$, i.e., $\beta_p$ being (at least) a two-fold zero. Thus, $\beta_p$ is a common zero of $\beta^p f(E,\beta)$ and $\partial_\beta (\beta^p f(E,\beta))$. Solving the resultant $f_\text{Res}\equiv\text{Res}_\beta[\beta^p f(E,\beta),\partial_\beta (\beta^p f(E,\beta))]=0$ yields all $E$ values where $f(E,\beta)$ possesses (at least) two-fold degenerate $\beta$-solutions. A subsequent step is to check which $E$ values indeed satisfy $\beta_p=\beta_{p+1}$ after sorting all the $\beta$-solutions in ascending order of modulus. For the Type-II and Type-III graph transitions discussed in the main text, two endpoints coincide on the OBC spectra exactly at the transitions. It implies that $f_\text{Res}$ as a function of $E$, should have a two-fold $E$-solution. Note that after removing the variable $\beta$, $f_\text{Res}$ depends only on $E$ and some tuning parameters (driving the transition). Hence there exists a common $E$-solution for $f_\text{Res}$ and $\partial_E f_\text{Res}$,  and the spectral transition is determined by $\text{Res}_E[f_\text{Res},\partial_E f_\text{Res}]=0$. Still, one has to check whether the coinciding endpoints indeed reside on the spectral graph.

\section{(IV) Calculations of isolated topological edge modes outside the spectral graph. }
In the main text, we have discussed how to account for isolated topological edge modes through the matrix $M_{edge}$. Formally, for a given $m$-band ($m\geq 2$) Hamiltonian $H(\beta)=\sum_{j=-s}^{t}h_j \beta^j$ (with $h_j$ being $m\times m$ matrices) and its associated ChP $f(E,\beta)=\text{det}(H(\beta)-E)=\sum_{j=-p}^{q}f_j \beta^j$, the isolated edge modes can be determined by solving:
\begin{align}
	\text{det}M_{edge}=0.
	\label{eqSM:detMedge}
\end{align}
The $\mu$th row and $\nu$-th column ($\mu,\nu=1,2,\cdots,s$) of the $ms\times ms$ matrix $M_{edge}$ is an $m\times m$ block, given by
\begin{eqnarray}
	[M_{edge}]_{\mu\nu}=\frac{1}{2\pi i}\int_{\Gamma} \beta^{\mu-\nu}(H(\beta)-E~I_{m\times m})^{-1} \frac{d\beta}{\beta}.
	\label{eqSM:Medge}
\end{eqnarray}
Here $I_{m\times m}$ is the identity matrix. The line integral is defined entry-wise, tracing a counterclockwise closed curve $\Gamma$ that encloses the origin and the first $p$ $\beta$-solutions of $f(E,\beta)$ with the smallest moduli. In the following, we apply the above formula to solve the isolated edge modes for two models: one with zero edge modes (the non-Hermitian SSH chain) and the other one with nonzero complex edge modes.

\subsection{(A) Non-Hermitian SSH model}
The non-Hermitian SSH model \cite{nhse1} with asymmetric hoppings is described by Hamiltonian:
\begin{align*}
	H(\beta)=[t_1+\frac{t_2}{2}(\beta+\beta^{-1})]\sigma_x+[\frac{t_2}{2i}(\beta-\beta^{-1})+i\frac{\gamma}{2}]\sigma_y,
\end{align*}
where $\sigma_{x,y}$ are Pauli matrices and $\gamma$ is the strength of nonreciprocity. Previously, the isolated topological edge modes have been studied in the framework of non-Bloch band theory. A topological invariant known as winding number defined on the GBZ \cite{nhse1} is utilized to characterize the edge modes. Here instead, we start with the ChP: 
\begin{subequations}
	\begin{align}
		f(E,\beta)=\text{det}\, \big(H(\beta)-E I_{2\times2}\big) 
		&=E^2-(t_1-\frac{\gamma}{2}+t_2 \beta)(t_1+\frac{\gamma}{2}+t_2 \beta^{-1}) \label{eqSM:det1} \\
		&:=-t_2(t_1+\frac{\gamma}{2})\beta^{-1}(\beta-\beta_1)(\beta-\beta_2). \label{eqSM:det2}
	\end{align}
\end{subequations}
The two roots of the ChP satisfy $|\beta_1|\neq|\beta_2|$ if $E\notin G_H$. $s=p=1$. $M_{edge}(E)$ is a $2\times 2$ matrix: 
\begin{align}
	M_{edge}(E)=\frac{1}{2\pi i}\int_{\Gamma}\frac{d\beta}{t_2(t_1+\gamma/2)(\beta-\beta_1)(\beta-\beta_2)}
	\begin{pmatrix}
		E & t_1+\gamma/2+t_2 \beta^{-1} \\
		t_1-\gamma/2+t_2 \beta & E
	\end{pmatrix}.
	\label{eqSM:MedgeSSH}
\end{align}
The path $\Gamma$ encloses the origin and the root of smaller modulus. From Eq. (\ref{eqSM:det1}), we have that if $E=0$, then $t_1+\gamma/2+t_2 \beta^{-1}=0$ for $\beta_1$ and $t_1-\gamma/2+t_2 z=0$ for $\beta_2$. In this case, the off-diagonal entries of the integrand in Eq. (\ref{eqSM:MedgeSSH}) will cancel out one of the singularities ($\beta_1$ or $\beta_2$ in the denominator). Otherwise, both singularities $\beta_1$ and $\beta_2$ will remain. Thus, we need to consider the $E=0$ and $E\neq0$ cases respectively.

\textit{\textbf{Case I:}} $E\neq0$. For this case,
\begin{align}
	(t_1+\gamma/2+t_2 \beta_j^{-1})(t_1-\gamma/2+t_2 \beta_j)\neq 0,\quad j=1,2.
	\label{eqSM:inequality}
\end{align}
To complete the integral via residue theorem, we assume $|\beta_1|<|\beta_2|$ and thus $\Gamma$ encloses the origin and $\beta_1$. Note that compared to $[M_{edge}(E)]_{21}$, $[M_{edge}(E)]_{12}$ has an extra singularity at the origin. Hence
\begin{align*}
	&[M_{edge}(E)]_{11}=[M_{edge}(E)]_{22}=\frac{E}{\beta_1-\beta_2}, \\
	&[M_{edge}(E)]_{12}=\frac{t_2}{\beta_1 \beta_2}+\frac{(t_1+\gamma/2)+t_2 \beta_1^{-1}}{\beta_1-\beta_2}, \\
	&[M_{edge}(E)]_{21}=\frac{(t_1-\gamma/2)+t_2 \beta_1}{\beta_1-\beta_2},
\end{align*}
where we have dropped the common constant factor $1/t_2(t_1+\gamma/2)$ for simplicity. Consequently,
\begin{align}
	\text{det}\, M_{edge}(E)&=[M_{edge}(E)]_{11} [M_{edge}(E)]_{22}-[M_{edge}(E)]_{12} [M_{edge}(E)]_{21}, \nonumber\\
	&=\frac{E^2}{(\beta_1-\beta_2)^2}-\frac{t_2}{\beta_1 \beta_2}\frac{(t_1-\gamma/2)+t_2 \beta_1}{\beta_1-\beta_2} -\frac{[(t_1-\gamma/2)+t_2 \beta_1][(t_1+\gamma/2)+t_2 \beta_1^{-1}]}{(\beta_1-\beta_2)^2}, \nonumber\\
	&=-\frac{t_2}{\beta_1 \beta_2}\frac{(t_1-\gamma/2)+t_2 \beta_1}{\beta_1-\beta_2}.
\end{align}
The first and third terms in the second row cancel each other, c.f., Eq. (\ref{eqSM:det1}). From Eq. (\ref{eqSM:inequality}), we have that for any $E\neq 0$, $\text{det}~M_{edge}(E)\neq 0$. That is, any $E\neq 0$ cannot be an edge mode.

\textit{\textbf{Case II:}} $E=0$. The two $\beta$-solutions are
\begin{align}
	\beta_1=-\frac{t_2}{t_1+\gamma/2},\quad \beta_2=-\frac{t_1-\gamma/2}{t_2}.
	\label{eqSM:roots}
\end{align}
And $M_{edge}(E)$ reduces to
\begin{align}
	M_{edge}(E)=\frac{1}{2\pi i}\int_{\Gamma}\frac{d\beta}{t_2(t_1+\gamma/2)(\beta-\beta_1)(\beta-\beta_2)}
	\begin{pmatrix}
		0 & (t_1+\gamma/2)\beta^{-1}(\beta-\beta_1) \\
		t_2(\beta-\beta_2) & 0
	\end{pmatrix}.
\end{align}
Obviously, $[M_{edge}(E)]_{11}=[M_{edge}(E)]_{22}=0$. And we have 
\begin{align}
	\begin{cases}
		\text{if}\; |\beta_1|<|\beta_2|,& [M_{edge}(E)]_{12}=-1/t_2 \beta_2,\\ &[M_{edge}(E)]_{21}=1/(t_1+\gamma/2), \\
		\text{if}\; |\beta_1|>|\beta_2|,& [M_{edge}(E)]_{12}= [M_{edge}(E)]_{21}=0.
	\end{cases}
\end{align}
Above we have used the fact that $\Gamma$ encloses the origin and $\beta_1$ ($\beta_2$) if $|\beta_1|<|\beta_2|$ ($|\beta_1|>|\beta_2|$). Thus, $E=0$ is the edge mode if and only if $|\beta_1|>|\beta_2|$ (so that det$M_{edge}(E)=0$), i.e., $|t_1^2-\gamma^2/4|<t_2^2$. It corresponds to the topological regime in Ref. \cite{nhse1}, as also verified by the numerical results in Fig. \ref{figS3}(a).

\subsection{(B) Topological model with non-zero edge modes}
We consider another model with Hamiltonian:
\begin{align}
	H(\beta)=t_1\sigma_0 + [u+\frac{v}{2}(\beta+\beta^{-1})]\sigma_x \; +\frac{v}{2i}(\beta-\beta^{-1})\sigma_y +i(g+t_2)\sigma_z.
	\label{eqSM:model2}
\end{align}
Compared to the nonreciprocal SSH model, this model breaks chiral symmetry due to the onsite complex potential. The ChP is
\begin{subequations}
	\begin{align}
		\text{det}\, \big(H(\beta)-E I_{2\times2}\big)
		&=[t_1+i(g+t_2)-E][t_1-i(g+t_2)-E] -(u+v\beta^{-1})(u+v\beta), \label{eqSM:detmodel2_1} \\
		&:=-uv\beta^{-1}(\beta-\beta_1)(\beta-\beta_2). \label{eqSM:detmodel2_2}
	\end{align}
\end{subequations}
$s=p=1$. $M_{edge}(E)$ is a $2\times2$ matrix:
\begin{align}
	&M_{edge}(E)=\frac{1}{2\pi i}\int_{\Gamma}\frac{d\beta}{uv(\beta-\beta_1)(\beta-\beta_2)} \begin{pmatrix}
		E-t_1+i(g+t_2) & u+v\beta^{-1} \\
		u+v\beta & E-t_1-i(g+t_2)
	\end{pmatrix},
	\label{eqSM:detMedgeModel2}
\end{align}
where $\Gamma$ encloses the origin and the $\beta$-solution of smaller modulus. Similar to the non-Hermitian SSH model, we discuss the $E\neq t_1\pm i(g+t_2)$ and $E= t_1\pm i(g+t_2)$ cases, respectively.

\textit{\textbf{Case I:}} $E\neq t_1\pm i(g+t_2)$. For this case,
\begin{align}
	(u+v\beta_j^{-1})(u+v\beta_j)\neq 0, \quad j=1,2.
	\label{eqSM:inequalityModel2}
\end{align}
Let us assume $|\beta_1|<|\beta_2|$. By the residue theorem, we have
\begin{align}
	&[M_{edge}(E)]_{11}=\frac{1}{uv(\beta_1-\beta_2)}[E-t_1+i(g+t_2)], \nonumber\\
	&[M_{edge}(E)]_{22}=\frac{1}{uv(\beta_1-\beta_2)}[E-t_1-i(g+t_2)], \nonumber\\
	&[M_{edge}(E)]_{12}=\frac{v}{uv \beta_1 \beta_2}+\frac{u+v\beta_1^{-1}}{uv(\beta_1-\beta_2)}, \nonumber\\
	&[M_{edge}(E)]_{21}=\frac{u+v\beta_1}{uv(\beta_1-\beta_2)}, \nonumber
\end{align}
and consequently 
\begin{align*}
	\text{det}\; M_{edge}(E)&=-\frac{v(u+v\beta_1)}{(uv)^2 \beta_1 \beta_2(\beta_1-\beta_2)} +[\frac{1}{uv(\beta_1-\beta_2)}]^2\times \\
	&\qquad\Big\{ [E-t_1+i(g+t_2)][E-t_1-i(g+t_2)] -(u+v\beta_1^{-1})(u+v\beta_1) \Big\}, \nonumber\\
	&=-\frac{v(u+v\beta_1)}{(uv)^2 \beta_1 \beta_2(\beta_1-\beta_2)}.
\end{align*}
Here the term in the curly bracket vanishes due to det$(H(\beta_1)-E I_{2\times2})=0$, cf. Eq. (\ref{eqSM:detmodel2_1}). From Eq. (\ref{eqSM:inequalityModel2}), we have det$M_{edge}(E)\neq 0$. That is, any $E\neq t_1\pm i(g+t_2)$ cannot be an edge mode.

\begin{figure}
	\includegraphics[width=0.6\textwidth]{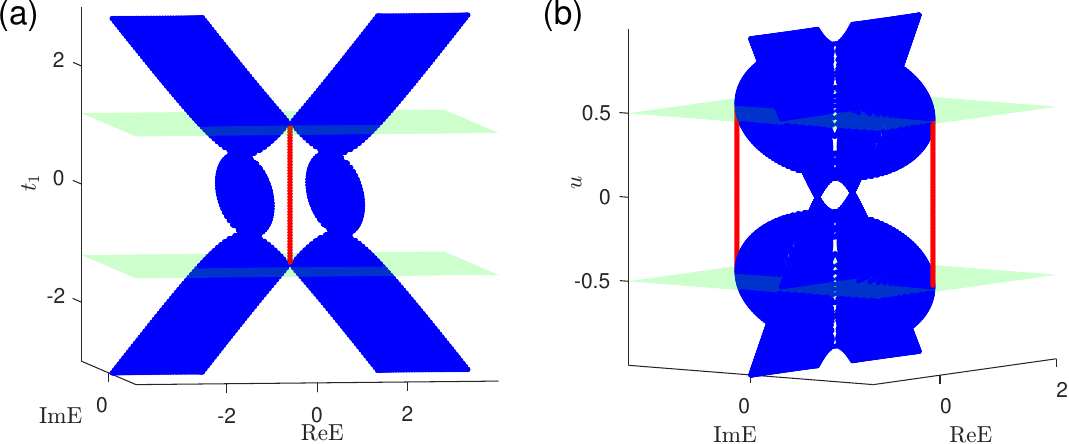}
	\caption{\label{figS3} Energy spectra under OBC for (a) the non-Hermitian SSH model with varying parameter $t_1$ and (b) the model Eq. (\ref{eqSM:model2}) with varying parameter $u$. The edge modes are highlighted in red. In (a), $t_2=1, \gamma=4/3$. The light green planes mark the topological phase transitions when $t_1=\pm\sqrt{t_2^2+(\gamma/2)^2}\approx \pm 1.2$. In (b), $t_1=0.3, t_2=0.1, g=0.3, v=0.5$. The phase transition occurs when $u=\pm v$.}
\end{figure}

\textit{\textbf{Case II:}} $E=t_1\pm i(g+t_2)$. For this case, the two $\beta$-solutions are
\begin{align}
	\beta_1=-\frac{v}{u}, \quad \beta_2=-\frac{u}{v},
	\label{eqSM:rootsModel2}
\end{align}
and Eq. (\ref{eqSM:detMedgeModel2}) reduces to 
\begin{align}
	M_{edge}(E)&=\frac{1}{2\pi i}\int_{\Gamma}\frac{d\beta}{uv(\beta-\beta_1)(\beta-\beta_2)} \begin{pmatrix}
		0 & u\beta^{-1}(\beta-\beta_1) \\
		v(\beta-\beta_2) & 0
	\end{pmatrix}.
\end{align}
It is straightforward to complete the integral via the residue theorem. $[M_{edge}(E)]_{11}=[M_{edge}(E)]_{22}=0$, and 
\begin{align}
	\begin{cases}
		\text{if}\; |\beta_1|<|\beta_2|, &\; [M_{edge}(E)]_{12}=-1/v\beta_2, \;[M_{edge}(E)]_{21}=1/u, \\
		\text{if}\; |\beta_1|>|\beta_2|, &\; [M_{edge}(E)]_{12}=[M_{edge}(E)]_{21}=0.
	\end{cases}
\end{align}
Thus det$M_{edge}(E)=0$ if and only if $|\beta_1|>|\beta_2|$, or $|u|<|v|$.  In this regime, the OBC spectra support topological edge modes with energies $E=t_1\pm i(g+t_2)$, as shown in Fig.\ref{figS3}(b).

\end{document}